\documentstyle{mn}

\begin{document}
\def\begeq{\begin{equation}}
\def\endeq{\end{equation}}
\def\begeqar{\begin{eqnarray}}
\def\endeqar{\end{eqnarray}}
\def\ac{accretion}
\def\mf{magnetic field}
\def\MHD{magneto\-hydro\-dyna\-mic}
\def\Amp{Amp\`ere}
\def\elm{electro\-magne\-tic}
\def\magn{mag\-ne\-tic}
\def\dy{dynamo}
\def\gm{gra\-vi\-to\-\magn}
\def\po{po\-lo\-idal}
\def\tor{to\-ro\-idal}
\def\Gauss{\rm Gauss}
\def\s{\rm s}
\def\cm{\rm cm}
\def\Msol{M_\odot}
\def\BL{Boyer\--Lindquist}
\def\equ{equation}
\def\ind{induction}
\def\indeq{\ind\ \equ}
\def\1d#1{{1\over#1}}
\def\del{\partial}
\def\delt#1{{\del#1 \over \del t}}
\def\ddtau#1{{d#1 \over d\tau}}
\def\delthe#1{{\del#1 \over \del\theta}}
\def\delr#1{{\del#1 \over \del r}}
\def\ti{\tilde}
\def\o{\omega}
\def\Om{\Omega}
\def\ot{\tilde \omega}
\def\a{\alpha}
\def\ag{\alpha_{\rm g}}
\def\rg{r_{\rm g}}
\def\vbeta{\vec{\beta}}
\def\phid{\hat\phi}
\def\E{\vec{E}}
\def\Ep{\vec{E}_{\rm p}}
\def\Eph{E^{\phi}}
\def\Ephid{E^{\phid}}
\def\B{\vec{B}}
\def\Bp{\vec{B}_{\rm p}}
\def\Bph{B^{\phi}}
\def\jph{j^{\phi}}
\def\jphid{j^{\phid}}
\def\j{\vec{j}}
\def\jp{\vec{j}_{\rm p}}
\def\v{\vec{v}}
\def\vt{\vec{v}_{\rm t}}
\def\vp{\vec{v}_{\rm p}}
\def\S{\vec{S}}
\def\ephi{\vec{e}_{\phi}}
\def\ephid{\vec{e}_{\hat\phi}}
\def\ethe{\vec{e}_{\theta}}
\def\ethed{\vec{e}_{\thed}}
\def\erd{\vec{e}_{\rd}}
\def\er{\vec{e}_r}
\def\ethe{\vec{e}_{\theta}}
\def\e{\vec{e}}
\def\vph{v^{\phi}}
\def\vrd{v^{\rd}}
\def\vthed{v^{\thed}}
\def\vphid{v^{\phid}}
\def\nonu{\nonumber}
\def\n{\vec{\nabla}}
\def\rot{\n \times}
\def\div{\n\cdot}
\def\tarrow#1{\buildrel\leftrightarrow\over #1}
\def\tens#1{\ifmmode\mathchoice{\mbox{$\sf\displaystyle#1$}}
{\mbox{$\sf\textstyle#1$}}
{\mbox{$\sf\scriptstyle#1$}}
{\mbox{$\sf\scriptscriptstyle#1$}}\else
\hbox{$\sf\textstyle#1$}\fi}
\newcommand{\subi}{_{\rm i}}
\newcommand{\sube}{_{\rm e}}
\newcommand{\mi}{m_{\rm i}}
\newcommand{\me}{m_{\rm e}}
\newcommand{\rhomi}{\rho_{\rm mi}}
\newcommand{\rhome}{\rho_{\rm me}}
\newcommand{\rhom}{\rho_{\rm m}}
\newcommand{\rhoci}{\rho_{\rm ci}}
\newcommand{\rhoce}{\rho_{\rm ce}}
\newcommand{\rhoc}{\rho_{\rm c}}
\newcommand{\gammai}{\gamma_{\rm i}}
\newcommand{\gammae}{\gamma_{\rm e}}

\title{Generation of Magnetic Fields by a Gravitomagnetic Plasma Battery} 
\author[Ramon Khanna]{Ramon Khanna\\
Landessternwarte, K\"onigstuhl, D-69117 Heidelberg, Germany\\
rkhanna@lsw.uni-heidelberg.de}

\maketitle

\begin{abstract}
The generation of magnetic fields by a battery, operating 
in an ion-electron plasma around a Kerr black hole, 
is studied in the 3+1 split of the Kerr metric. 
It is found that the gravitomagnetic contributions to the electron 
partial pressure are able to drive currents.
The strength of the equilibrium magnetic field should be higher than
for the classical Biermann battery, which is found to operate in this 
relativistic context as well, since the gravitomagnetic driving terms can 
less easily be quenched than the classical ones.
In axisymmetry the battery can induce only toroidal magnetic fields. 
Once a toroidal magnetic field is present, however, the coupling of 
gravitomagnetic and electromagnetic fields generates a poloidal magnetic 
field even in axisymmetry (Khanna \& Camenzind 1996a,b).
A rotating black hole, embedded in plasma, will therefore always 
generate toroidal and poloidal magnetic fields.
\end{abstract}


\section{Introduction}
In the year 1950 Ludwig Biermann published a paper `On the Origin of Magnetic 
Fields on Stars and in Interstellar Space'. He showed that currents must flow 
in a plasma, if its effective acceleration possesses a rotational part and thus 
charge separation due to the electron partial pressure can not be balanced 
by an electrostatic field. In the case of an initially 
non-magnetized, differentially rotating star with no meridional circulation, 
this battery would generate \mf s of the order kGauss. 
Mestel \& Roxburgh (1962) revisited `Biermann's battery', considering the 
effects of meridional circulation and the presence of a weak \po\ \mf . 
They concluded that,
due to Lorentz forces modifying the plasma rotation, the \mf\ will saturate at 
much lower field strength than it would due to Ohmic losses. As a 
consequence Biermann's battery could no longer be considered as driving 
mechanism behind the strong \mf s of rotating stars, but could only 
provide seed fields for a \dy .

Recently Biermann's battery has experienced a renaissance as provider of 
\mf s in the central tori of active galactic nuclei. The joint action 
of the rapid (Keplerian) differential rotation of these objects with 
fast outflow motions (and in some cases \dy\ action) was considered an 
attractive mechanism for the production and spreading of \mf s in galaxies 
(e.g. Lesch et al. 1989; Chakrabarti 1991; Chakrabarti et al. 1994; 
Andreyasan 1996).

`Batteries' driven by gravitational-electromagnetic coupling have also been 
studied in the literature. Thorne et al. (1986) considered the model problem of 
a rotating black hole in vacuum, threaded by a stationary electric field. The 
coupling between gravitomagnetic potential of the hole and the electric field 
requires a stationary \mf . The electric field was, however, assumed to be due 
to a charged black hole, which is probably not very relevant in the 
astrophysical context. 
Opher \& Wichoski (1997) showed that nonminimal gravitational-electromagnetic 
coupling in protogalaxies could generate \mf s of $\sim 10^{-6} \Gauss $. 

In the present paper I combine both Biermann-type battery effects with 
gravitomagnetic effects. The classical derivations
of Biermann (1950) and Mestel \& Roxburgh (1962) are repeated using a 
new set of equations governing an inviscid (but resistive) plasma in the 
vicinity of a
Kerr black hole, which have been derived by Khanna (1998) within 
the 3+1 split of the Kerr metric. This set of \equ s contains a generalized 
Ohm's law, which is fundamental for the study of battery effects. 
I obtain a general relativistic expression for the `Impressed Electric 
Field' (IEF), which is due to the electron partial pressure in a plasma in the 
vicinity of a rotating black hole. I show that, in axisymmetry, the \gm\ 
contributions to the IEF are not curl-free and therefore drive currents 
and generate \mf s. The total IEF is likely to be rotational in general.

The metric has signs \((-{}+{}+{}+)\) and $c=1$. 
Vectors and tensors in 3-dimensional space are denoted by arrows 
(e.g. $\v$ and $\tarrow{T}$) and covariant derivatives are indicated by 
vertical bars, e.g. \(\nabla_j \beta^i \equiv \beta^i_{\; |j}\).

\section{A plasma battery in the Kerr metric}
In this section I will first review the 3+1 split of the Kerr metric.
Secondly, the \equ\ of motion and the generalized Ohm's law 
will be combined to find an expression for the IEF that drives currents in a 
quasi-neutral plasma (quasi-neutral in the plasma rest frame) surrounding 
a rotating black hole.

\subsection{The 3+1 split of the Kerr metric}
In the 3+1 split of the Kerr metric (Thorne et al. 1986) spacetime is split 
into a family of three dimensional differentially rotating 
hypersurfaces of constant time with internal curvature. 
These hypersurfaces of constant time can be mentally collapsed 
into a single 3-dimensional `absolute space' in which time is {\it globally}
measured by the \BL\ coordinate $t$. Physics is described in absolute space by 
locally non-rotating fiducial observers (FIDOs) with respect to their {\it 
local} proper time $\tau$ in their locally flat frames.
The line element of the Kerr metric in 3+1 notation is given by
\begeqar
        ds^2 &=& g_{\a\beta}dx^{\a}dx^{\beta} \nonu\\
	&=& 
	-\ag^2\, dt^2 + h_{jk}(dx^j + \beta^j\, dt)
                                         (dx^k + \beta^k\, dt)  \; ,
\label{dsKerr31}
\endeqar
where the lapse function is identified with the gravitional redshift
\begeq
        \ag\equiv (d\tau /dt)_{\rm FIDO} = {\rho\over\Sigma}\sqrt{\Delta}\; ,
\endeq
the shift functions are the components of the gravitomagnetic potential 
$\vbeta$ with
\begeq
        \beta^r = \beta^{\theta} = 0\; ,\quad
            \beta^{\phi}\equiv -\o = -(d\phi / dt)_{\rm FIDO} = 
	- {2aMr\over\Sigma^2}\; ,
\label{shift}
\endeq
which describes the differential rotation of the FIDOs relative to distant 
observers. The components of the 3-metric $\tarrow{h}$ are 
\begeq
        h_{rr} = {\rho^2\over \Delta}\; ,\quad
            h_{\theta\theta} = \rho^2\; ,\quad
            h_{\phi\phi} = \ot^2\; ,\quad
            h_{jk} = 0\ {\rm if}\ j\ne k                     \; .
\label{hikKerr}
\endeq
Note that $h_{jk} = g_{jk}$ but $h^{jk}= g^{jk}+\beta^j\beta^k / \ag^2$.
The metric functions appearing here are defined as
\begeqar
        \Delta &\equiv & r^2-2Mr + a^2\; ,\quad
            \rho^2 \equiv r^2+a^2\cos^2{\theta}\; , \nonu\\
        \Sigma^2 &\equiv & (r^2+a^2)^2-a^2\Delta\sin^2\theta\; ,\quad
            \ot \equiv (\Sigma /\rho )\sin\theta \; .
\endeqar
The parameters of the black hole are its mass $M$ and angular momentum 
$J$, which define the Kerr parameter \(a\equiv J/M\). 

The FIDO-measured velocities are
\begeq
	v^j = \1d{\ag}\left(\frac{dx^j}{dt} + \beta^j\right)\, ,\hbox{ e.g. }
	\vph = \frac{\Om - \o}{\ag}\; .
\endeq
The physical velocity components $v^{\hat j}$ follow by multiplcation with 
$(h_{jj})^{1/2}$.

\subsection{Battery theory}
In a non-magnetized plasma, currents are driven by electric fields and 
the `impressed electric field'. If the IEF possesses a potential, 
it may be cancelled by an electrostatic field and no current will flow.
The IEF is included in a generalized Ohm's law, which is the base of 
battery theory. 

If the plasma is `cold' (pressure and internal energy of the components $\ll$ 
mass energy density) and quasi-neutral in its rest frame the generalized 
Ohm's law in the MHD limit is given by (Khanna 1998)
\begeq
        \j = 
	\sigma\gamma(\E +\v\times\B) -\frac{\sigma}{e n}(\j\times\B) 
	+\frac{\sigma}{e n\ag}\n(\ag p\sube)\; .
\label{allgOhmqn}
\endeq
Here $\j$, $\E$, $\B$, $\v$ and $\gamma$ have their usual meaning, but are 
measured locally by FIDOs. $n$, $p\sube$ and $\sigma$ are electron density,
electron pressure and plasma conductivity, respectively, defined in the 
plasma rest frame (note that I do 
not distinguish between electron, ion and plasma rest frame. This is because 
treating a plasma as single fluid implies that electrons and ions have 
non-relativistic bulk velocities in the plasma rest frame; Khanna 1998). 
The `impressed electric field'
\begeq
	\E^{(i)} = \frac{\n(\ag p\sube)}{e n\gamma\ag}
\label{IEF}
\endeq
can be re-expressed with the aid of the \equ\ of motion to see, if it possesses 
a potential or not. To be correct, it is \(\ag\E^{(i)}\) that matters, since 
the curl of \(\ag\E\) appears in Faraday's law (see below).

The \equ\ of motion for a `cold' and quasi-neutral plasma 
(then the charge density measured by FIDOs is \(\rhoc = \v\cdot\j\)) has 
the form (Khanna 1998)
\begeqar
        \rhom\gamma\ddtau{(\gamma\v)}&=&
        \rhom\gamma^2\vec{g} + \rhom\gamma^2\tarrow{H}\cdot\v 
	- \1d{\ag}\n(\ag p)\nonu\\
	&+& (\v\cdot\j)\E + \j\times\B \; ,
\label{stateqmot}
\endeqar
with the energy density of the plasma defined in the plasma rest frame
\(\rhom\approx n\subi m\subi\) and $n\subi$ is the ion density.
The total time derivative in the FIDO frame is defined as 
\begeq
	\ddtau{} \equiv \1d{\ag}\left[\delt{} + (\ag\v-\vbeta)\cdot\n\right]\; .
\endeq
The total plasma pressure is given by
\begeqar
	p &\equiv& p\subi + p\sube = n\subi k T + \left(n\sube k T 
	+ \frac{b}{3} T^4\right) 
	\nonu\\
	&=& {}\left(\frac{n\subi}{n\sube} + 1\right)p\sube
	-\frac{n\subi}{n\sube}\frac{b}{3} T^4\; .
\label{totpress}
\endeqar
Note that the radiation pressure appears as part of the electron partial 
pressure. This is because it is assumed that the interaction between radiation 
field and plasma is dominated by Compton scattering. Secondly 
\(p \equiv p\subi + p\sube \) is a 
definition from the two-component plasma theory, which the \equ s used here 
are based upon. Therefore photons are not treated as a third component (effects
of the radiation pressure will not be discussed in detail anyway).
For a quasi-neutral plasma (\(Z n\subi\approx n\sube \equiv n\)) 
Eq.~(\ref{totpress}) becomes
\begeq
        p\sube \approx \frac {Z}{Z+1}\left(p 
		+ \frac{b}{3 Z} T^4\right)
\endeq
and the IEF is given by
\begeq
        \E^{(i)} \approx \frac{Z}{e n \gamma\ag(Z+1)}\left(\n(\ag p)
		+ \frac{b}{3Z}\n(\ag T^4)\right)\; . 
\endeq
Eliminating \(\n(\ag p)/\ag\) with the aid of Eq.~(\ref{stateqmot}) yields
\begeqar
	\lefteqn{ \E^{(i)} \approx 
	\frac{m\subi}{(Z+1)e }
	\left(\gamma\vec{g} + \tarrow{H}\cdot(\gamma\v) 
        -\ddtau{(\gamma\v)}\right)
	}\nonu\\
	&&{}+\frac{Z \left(\j\times\B + (\v\cdot\j)\E\right) }
		{(Z+1) e n \gamma}\, 
	+ \frac{b\n(\ag T^4)}{3 (Z+1) e n \gamma\ag} \; .
\endeqar
This expression shows that, for relativistic plasma velocities, even gravity 
will drive currents, unless $\gamma$ is a function of $\ag$ only (since 
\(\vec{g} = -\n\ln\ag\)) or a constant. For most practical purposes, however, 
$\gamma\approx 1 $, even close (but not too 
close) to a rotating black hole, and the IEF due to gravity will be balanced 
by an electrostatic field. I note in passing that the IEF due to radiation 
pressure will also be rotational unless $\gamma n$ is a function of the 
radiation pressure. 
This topic will, however, not be pursued any further in this paper.

Inserting $\E^{(i)}$ into Ohm's law yields
\begeqar
	\j&\approx&\sigma\gamma\left(
		\left(\frac{Z (\v\cdot\j) }{(Z+1) e n \gamma} + 1\right)\E
		+ \v\times\B\right) 	\nonu\\
	&+&\frac{\sigma \gamma m\subi}{(Z+1) e}\left(
			\gamma\vec{g} + \tarrow{H}\cdot(\gamma\v) 
    -\ddtau {(\gamma\v)}\right)
	\nonu\\
	&-& \frac{\sigma}{(Z+1) e n}\j\times\B 
	 + \frac{\sigma b \n(\ag T^4)}{3(Z+1) e n\ag}  \; .
\label{Ohmmot}
\endeqar
The \mf\ associated with the current described by Eq.~(\ref{Ohmmot}) is given 
by \Amp 's law
\begeq
	\rot{(\ag\B)} = \left(\delt{} - {\cal L}_{\vbeta}\right)\E + 4\pi\ag\j
	\; .
\label{Amp}
\endeq
Note that the sources of magnetic fields are not only currents 
and the displacement current, but also the coupling between the 
gravitomagnetic potential and electric fields.
This has interesting consequences, which show up as `battery 
effects' (Thorne et al. 1986) or \dy\ effects (Khanna \& Camenzind 1996a,b). 
The generation of \mf s, however, requires, as usual, a 
rotational `renormalized' electric field, as demanded by Faraday's law 
\begeq
        \rot{(\ag\E)} = -\left(\delt{} - {\cal L}_{\vbeta}\right)\B \; .
\endeq
The Lie-derivative is defined as, e.g. 
\({\cal L}_{\vbeta}\B \equiv (\vbeta\cdot\n)\B - (\B\cdot\n)\vbeta\). 
It is straight forward to show that, since \(\div{\B}=0\),  
\({\cal L}_{\vbeta}\B = - \rot{(\vbeta\times\B)}\).
Faraday's law can therefore be written as 
\begeq
        \rot{(\ag\E + \vbeta\times\B)} = -\delt{\B}\; .
\label{Fara2}
\endeq
This means that stationary \mf\ requires \(\ag\E + \vbeta\times\B = -\n\Phi\), 
where $\Phi$ is a scalar potential.
Maxwell's \equ s are completed by
\begeq
	\div{\B} = 0 \quad{\rm and}\quad\div{\E} = 4\pi\rhoc.
\endeq

\subsubsection{The \gm\ battery}
Let us, for the moment, restrict ourselves to the {\it generation} of \mf\ by 
the electron partial pressure, i.e. let us drop the terms containing $\B$ 
and neglect the electric force term $(\v\cdot\j)\E$ in Eq.~(\ref{Ohmmot}). 
Leaving aside 
contributions from $\gamma\vec{g}$ and from radiation pressure, currents 
will be driven by the rotational parts of $\E$, the relativistic equivalent of 
the classical IEF
\begeq
	\E^{(i)}_{\rm class}= -\frac{m\subi}{(Z+1) e} 
	\left(\v\cdot\n\right)(\gamma\v)\; , 
\endeq
which is responsible for Biermann-type batteries, the \gm\ IEF
\begeq
	\E^{(i)}_{\rm gm}=\frac{m\subi}{(Z+1) e\ag} 
	\left(\vbeta\cdot\n + \n\vbeta\, \cdot\right)(\gamma\v)
\endeq
and the IEF due to time-dependent plasma motion
\begeq
	\E^{(i)}_{\del t} = -\frac{m\subi}{(Z+1) e\ag}\delt{(\gamma\v)}\; .
\endeq
$\ag\E^{(i)}_{\del t}$ will contribute to the current, if 
\(\rot{(\gamma\v)}\neq 0\), i.e. if there are non-stationary vortices in the 
fluid.

In the following I will analyse $\ag\E^{(i)}_{\rm class}$ and 
$\ag\E^{(i)}_{\rm gm}$ for \(\v = v^r \er + v^{\phi} \ephi\) and discuss under 
which circumstances those IEFs possess a potential: From 
$\ag\E^{(i)}_{\rm class}$ we have
\begeqar
   \lefteqn{-\left(\ag\v\cdot\n \right)(\gamma\v) 
	=-(\ag v^i)(\gamma v^j)_{|i}\, \e_j	}	\nonu\\
   &&=\ag\gamma(\vph)^2\ot \n\ot			\nonu\\
   &&-\; \ag\left(v^r(\gamma v^r)_{,r} + \gamma (v^r)^2(\ln\sqrt{h_{rr}})_{,r} 
			+  v^{\phi}(\gamma v^r)_{,\phi}\right) \er \nonu\\
   && +\; \ag\gamma (v^r)^2 (\ln\sqrt{h_{rr}})_{,\theta} \ethe \nonu\\
   && -\; \ag \left( v^r (\gamma v^{\phi})_{,r} 
	+ 2\gamma v^{\phi} v^r (\ln\ot)_{,r} 
	+ v^{\phi}(\gamma v^{\phi})_{,\phi}\right)\ephi \; .
\label{Eclgm}
\endeqar
Though it is unlikely that this expression should be irrotational in general, 
it is not possible to make statements, unless $v^r$ and $\vph$ are known 
explicitly. If, however, \(v^r = 0\) in an axisymmetric problem, 
only the first term on the r.h.s. remains. 
This is the relativistic equivalent of Biermann's IEF in a star. This 
expression is irrotational only if \(\ag\gamma (v^{\phi})^2\) is a function 
of $\ot$ alone, i.e. if, for large radii, \(\gamma (v^{\phi})^2\) is 
constant on cylinders. This is unlikely 
and certainly not possible within the ergosphere of the black hole; but there 
$v^r\neq 0$ anyway. 
The relativistic equivalent of 
Biermann's battery contributes therefore also around black holes.
The function part of \(\ag\E^{(i)}_{\rm gm}\) is
\begeqar
	\lefteqn{
	\left(\vbeta\cdot\n + \n\vbeta\, \cdot\right)(\gamma\v) = 
	\left(\beta^i(\gamma v^j)_{|i} + \gamma\beta^{i|j} v_i\right) \e_j 
	}\nonu\\
	&&= -\gamma\vph\ot^2\n\o 
            -\o\left((\gamma v^r)_{,\phi}\er
		+ (\gamma v^{\phi})_{,\phi}\ephi\right) \; .
\label{Egm}
\endeqar
In axisymmetry \(\ag\E^{(i)}_{\rm gm}\) is clearly rotational, 
unless some freak $\gamma$ should manage to make \(\gamma\vph\ot^2\) a function 
of $\o$ alone. Only if $\vph$ is non-axisymmetric, the \gm\ IEF drives \tor\ 
currents. With Eqs.~(\ref{Eclgm}) and (\ref{Egm}), it seems 
obvious that $\ag(\E^{(i)}_{\del t}+\E^{(i)}_{\rm class}+\E^{(i)}_{\rm gm})$ 
will be rotational in general.
Note, however, that if axisymmetry is imposed the \tor\ component of the total 
IEF (Eq.~[\ref{IEF}]) is zero, which means that, even though 
\(\ag(\E^{(i)}_{\del t}+\E^{(i)}_{\rm class})\) 
may have a rotational \tor\ component, no \tor\ currents will be driven. 
In axisymmetry poloidal \mf s can not be generated by the IEF. Once a \tor\ 
\mf\ has been induced, however, the \gm\ term
${\cal L}_{\vbeta}\E$ will generate a \po\ \mf\ (Khanna \& Camenzind 1996a; 
and see below).

\subsubsection{The equilibrium \mf }
We can now be confident that, but for pathological velocity fields, the \gm\ 
battery will generate \mf s. 
The important question that remains to be answered is how 
strong these fields can become. In the case of Biermann's battery, Mestel and 
Roxburgh (1962) have shown that the fields are limited to very low field 
strength, and are only significant as seed fields for a dynamo. In a 
self-consistent picture the limitation of the \mf\ is not due to Ohmic 
dissipation, but occurs at much lower level due to the back reaction of the 
\mf\ on the velocity field. 

The evolution of the \mf\ generated by the battery is given by 
Eq.~({\ref{Fara2}).
Eliminating $\E$ with the aid of Eq.~(\ref{Ohmmot}) (neglecting rotational 
contributions from $\gamma\vec{g}$ and radiation pressure) and using 
Eq.~(\ref{Amp}) we get the {\it generalized} MHD \indeq\
\begeqar
	\lefteqn{
        \delt{\B} =
        \rot{\left(\left(\frac{\ag\v}{\, \kappa}-\vbeta\right)\times\B\right)}
	}\nonu\\
	&&-\rot{ \frac{\eta}{\gamma\, \kappa} \left(\rot{(\ag\B)}-
		\left(	\delt{}-{\cal L}_{\vbeta}  \right)\E\right)
		} \nonu\\
        &&-{}\rot{\frac{\ag}{(Z+1) e\, \kappa}\left(\frac{\j\times\B}{n\gamma} 
        + \mi\left( \ddtau{} -\tarrow{H}\cdot\right)(\gamma\v) \right) }
        \; , 
\label{genindeq}
\endeqar
where \(\kappa = 1+ Z(\v\cdot\j)/(Z+1)en\gamma\). 
The \magn\ diffusivity has been introduced 
as \(\eta = c^2 / (4\pi\sigma)\). This \equ\ merits some discussion: The first 
two lines (if $\kappa = 1$) are the MHD \indeq , which one derives with the 
standard 
relativistic Ohm's law. In special relativity there is no ${\cal L}_{\vbeta}\E$
and Cowling's theorem against axisymmetric \dy\ action (Cowling 1934) holds. 
The presence of ${\cal L}_{\vbeta}\E$ in the context of the Kerr metric states 
that Cowling's theorem does, in principle, not hold in the vicinity of a 
rotating black hole. In the axisymmetric standard-Ohm-case and if the 
displacement current is neglected, 
$\E$ can be eliminated and the \equ\ can be split into \po\ and \tor\ parts 
to yield two coupled partial differential \equ s with ${\cal L}_{\vbeta}\E$ 
(expressed by the \po\ and \tor\ \mf s)
appearing as source term for the \po\ \mf\ (Khanna \& Camenzind 1996a). 

In the general case, 
Eq.~(\ref{genindeq}) is not of much practical use, since $\E$ and $\j$ 
can not be eliminated. But it serves for discussion of the processes. The \equ\ 
describes the joint action of battery and \gm\ \dy . Although numerical 
simulations of the kinematic axisymmetric \gm\ \dy\ with standard Ohm's law 
have so far failed to produce growing or stationary modes (Khanna \& Camenzind 
1996b; but see also N\'u\~nez 1997), it may well be that they appear with the 
aid of the IEF.
 
In stars the limitation of the thermal battery is due to modification of 
$\v$ by Lorentz forces. This has a quadratic effect on 
$(\v\cdot\n)\v$ so that ultimately $\rot{\E}=0$ can be achieved.
The \gm\ battery terms are linear in $\v$ so that the approach 
of the IEF plus limiting terms to an irrotational state can be 
expected to take place much slower, which should result in a higher value of 
the asymptotic \mf . 

Quantitative analytic estimates of the asymptotic state in this 
even more complicated system than the one considered by Mestel \& Roxburgh 
(1962) for the Biermann battery are very hard to achieve and can only be very 
crude. The solution of the battery/dynamo problem in Kerr metric has to be 
postponed until a numerical treatment of MHD is feasible.

\section{Conclusion}
It has been shown that in a quasi-neutral MHD-plasma, which is in axisymmetric 
motion around a rotating black hole, the \gm\ `Impressed Electric Field' is 
rotational and will drive currents, thus generating a \mf . For the general 
case the \gm\ IEF is likely to be rotational, but explicit velocity fields will 
have to be studied. The same holds for the total IEF. Poloidal \mf s can only 
be induced, if the electron partial pressure is non-axisymmetric.

In Kerr metric the battery problem can not be treated separately from the 
\gm\ \dy\ problem. It is therefore very unlikely that reliable analytical 
estimates of the equilibrium field strength of the system can be achieved.
It is argued that, since the \gm\ field of the rotating black hole can 
not be quenched by Lorentz forces (braking of the hole's rotation by battery 
generated fields is negligible) and 
the \gm\ battery terms are only linearly dependent on the velocity field of 
the plasma, the equilibrium field should be stronger than in the case 
of the classical Biermann battery.

The expression for the total IEF reveals the fact that, for relativistic 
motions with spatially variable Lorentz factor, even gravity will make 
rotational contributions to the IEF, i.e. it would drive currents even in 
Schwarzschild metric. Also time-dependent vortices of the velocity field 
drive currents.

This work is based on the generalized Ohm's law for a quasi-neutral plasma 
in the MHD limit. In a charged plasma there are `acceleration currents' and 
`\gm\ currents', which might give qualitatively new results. On timescales of 
the electron collision time, `current acceleration' terms become important, 
which might even introduce additional axisymmetric dynamo effects 
(Khanna 1998).

Radiation pressure does also contribute to the rotational part of the IEF. 
This has not been discussed in detail, but could be
significant in the hot radiative region close to the horizon. In this region, 
the single-fluid approach of MHD might break down alltogether (Khanna 1998) 
and the conlusions drawn here would have to be checked in a multi-fluid model.

\section*{Acknowledgements}
I thank Max Camenzind, Jochen Peitz, Stefan Spindeldreher and Markus Thiele 
for fruitful discussions. This work is supported by the Deutsche
Forschungsgemeinschaft (SFB 328).

\end{document}